\documentclass[times,twocolumn, final]{elsarticle}

\usepackage{jasr}
\usepackage{framed,multirow}

\usepackage{amssymb}
\usepackage{latexsym}

\usepackage[switch]{lineno}

\usepackage{url}
\usepackage{xcolor}
\definecolor{newcolor}{rgb}{.8,.349,.1}
\usepackage[normalem]{ulem}

\usepackage[citebordercolor=white]{hyperref}
\usepackage{etoolbox}

\usepackage{amsmath}		
\usepackage{comment}
\usepackage[normalem]{ulem}
\usepackage{gensymb}

\colorlet{orange}{.}

\newif\ifhighlight
\highlightfalse

\newcommand{\citealtc}[1]{\citeauthor{#1}, \citeyear{#1}}

\journal{Advances in Space Research}

\begin{document}

\verso{Korolkov \& Izmodenov}

\begin{frontmatter}


\title{New possible way to determine stellar wind terminal velocity from analysis of Lyman-$\alpha$ absorption spectra}


\author[1]{Sergey D. \snm{Korolkov}\corref{cor1}}
\cortext[cor1]{Corresponding author at: Space Research Institute of the Russian Academy of Sciences, Profsoyuznaya Str. 84/32, Moscow 117997, Russia.\\{\it Email address:} sergey.korolkov@cosmos.ru (S.D. Korolkov).}

\author[1,2,3]{Vladislav V. \snm{Izmodenov}\corref{cor2}}
\cortext[cor2]{Space Research Institute of the Russian Academy of Sciences, Profsoyuznaya Str. 84/32, Moscow 117997, Russia.\\{\it Email address:} izmod@cosmos.ru (V.V. Izmodenov).}

\affiliation[1]{organization={Space Research Institute of the Russian Academy of Sciences},
                addressline={Profsoyuznaya Str. 84/32},
                city={Moscow},
                postcode={117997},
                country={Russia}}

\affiliation[2]{organization={Lomonosov Moscow State University},
                addressline={1 Leninskie Gory},
                city={Moscow},
                postcode={119991},
                country={Russia}}
                
\affiliation[3]{organization={HSE University},
                addressline={20 Myasnitskaya Ulitsa},
                city={Moscow},
                postcode={101000},
                country={Russia}}

\received{--- 2025}
\finalform{--- 2025}
\accepted{--- 2026}
\availableonline{--- 2026}

\begin{abstract}


Stellar winds interact with the surrounding interstellar medium, forming circumstellar envelopes known as astrospheres. When the interstellar medium is partially ionized (as in the Sun's case), a distinctive feature of this interaction, the so-called hydrogen wall, is formed. The hydrogen wall consists of secondary interstellar atoms created through charge exchange processes near the tangential discontinuity that separates the stellar wind from the ionised component of the interstellar wind. This secondary interstellar atomic component is decelerated and slightly heated compared to the primary component, which consists of the original, unperturbed interstellar hydrogen atoms. As a result, the hydrogen wall can be detected via Lyman-$\alpha$ absorption spectra toward nearby stars. Both the heliospheric hydrogen wall and analogous structures around other nearby stars have been observed using the Hubble Space Telescope.


In this paper, we propose that another distinctive feature of the interaction between stellar winds and a partially ionised interstellar medium can potentially be detected in the Lyman-$\alpha$ absorption spectra of nearby stars. This feature, the so-called neutral stellar wind, is formed by charge exchange between supersonic stellar wind protons and interstellar atoms that penetrate deep into the astrosphere due to their long mean free paths. We present a parametric numerical analysis of astrospheric structures and their corresponding synthetic Lyman-$\alpha$ absorption spectra. This is achieved by varying the terminal velocity of the stellar wind while maintaining a constant dynamic pressure, thereby ensuring the astrosphere's size remains consistent across models. Using a two-dimensional kinetic-hydrodynamic model, we demonstrate that for stellar winds slower than the solar wind (with terminal velocities $V_0 \lesssim 200$ km/s), the efficiency of charge exchange in the supersonic region increases dramatically. In such astrospheres, the neutral stellar wind produces a distinct and observable absorption feature in the Lyman-$\alpha$ spectrum. While this signature is negligible for solar-like winds ($V_0 \approx 400$ km/s), it emerges as a direct spectroscopic diagnostic for winds with velocities up to $\sim 200$ km/s. The detection of this neutral wind absorption provides a novel method to directly constrain stellar wind velocities.

\end{abstract}


\end{frontmatter}


\section{Introduction}


The interaction between the supersonic solar wind and the partially ionised local interstellar medium forms a region characterised by several key discontinuities: the termination shock, the heliopause -- a tangential discontinuity separating the solar and interstellar plasmas -- and, possibly, a bow shock \citep[e.g.,][]{Baranov1970, Izmodenov2000}. A fundamental process governing the structure and dynamics of this region is charge exchange between solar wind protons and interstellar neutral hydrogen atoms \citep[e.g.,][]{baranov1993}. This process results in the creation of secondary interstellar atoms -- another population of neutral hydrogen that has a smaller bulk velocity and is hotter as compared to primary pristine interstellar hydrogen. The number density of the secondary population increases in the vicinity of the heliopause, forming a so-called hydrogen wall.


The existence of the hydrogen wall around the heliopause was first theoretically proposed by \cite{baranov1991, baranov1993}. Analysing Ly $\alpha$ absorption spectrum toward $\alpha$ Centauri \cite{linsky_wood1996} have found that the spectrum is redshifted and has a higher temperature as compared with the other interstellar lines. They have found that the best way to resolve the discrepancy between hydrogen Ly-$\alpha$ and other lines is to introduce the second absorption component. The parameters of the secondary component that are obtained from the spectral fit correspond to the parameters of the secondary hydrogen component obtained by \cite{baranov1993, Baranov1995}. Therefore, the existence of the hydrogen wall around the heliopause has been observationally confirmed. This seminal work of \cite{linsky_wood1996} established Lyman-$\alpha$ absorption as a unique remote-sensing tool for probing the outer heliosphere. Later, \cite{Wood1998} reported the detection of analogous astrospheric hydrogen walls around other stars. These and subsequent detections of astrospheric absorption features have provided a revolutionary method for indirectly studying solar-like stellar winds \citep{Wood2002}. It should be noted that other indirect diagnostic methods also exist, a comprehensive review of which can be found, for example, in \citet{Vidotto2021}.



The primary value of this technique is its ability to constrain stellar mass-loss rates ($\dot{M}_\star$), a key parameter in stellar evolution. A series of studies led by Wood and collaborators derived $\dot{M}_\star$ for a sample of solar-like stars, finding that mass-loss correlates with magnetic activity and decreases with stellar age \citep{Wood2002, Wood2005ApJ}. Moreover, \cite{Wood2004} proposed the concept of a "wind dividing line" at an age of approximately 600 Myr, suggesting a sharp transition where mass-loss rates for solar-type stars increase dramatically, implying that younger Suns have even weaker winds. Thus, even in their youth, solar-type stars do not exhibit evidence for significantly stronger winds \citep{Wood2014}. In contrast, studies of evolved stars, such as the recent analysis of M-giants by \cite{Wood2024}, show that these stars can have much stronger and slower winds, with estimated $\dot{M}_\star$ on the order of $14-86 \times 10^{-11}\ M_{\odot}\ \mathrm{yr}^{-1}$ and velocities of 20-40 km/s. 

While early studies primarily identified Ly $\alpha$ absorption from the hydrogen wall in the upwind directions, evidence for absorption originating from the inner heliosheath has been pointed out by \cite{izmodenov1999}. A significant advance was made by \cite{Wood2007}, who systematically analysed reconstructed stellar Ly $\alpha$ profiles and found that the lines of sight with \textcolor{orange}{orientations greater than 160 degrees from the upwind direction} exhibited significant blueshifts. This was interpreted as the signature of broad, previously undetected absorption from neutrals in the heliosheath, a region where extended path lengths make such absorption dominant. This finding not only provided clear evidence for heliosheath Ly $\alpha$ absorption but also highlighted the challenges in detecting its characteristically broad spectral signature compared to the sharper absorption feature from the hydrogen wall.

The quantitative interpretation of Ly $\alpha$ spectra requires self-consistent numerical models of the plasma-neutral interaction. Pioneering work by \cite{baranov1993, Baranov1995} developed the first self-consistent kinetic-hydrodynamic model, treating the interstellar hydrogen kinetically via the Boltzmann equation to account for their large mean free paths. This "Baranov-Malama" model became the foundation for interpreting heliospheric and astrospheric data. Wood and collaborators later used these and other models (e.g., four-fluid models in \cite{Wood2001, Wood2002, Wood2005ApJ, Wood2014}; improved kinetic models in \cite{Izmodenov2002}) to match observed absorption and derive stellar wind properties.

\textcolor{orange}{In the pioneering studies by Wood and collaborators, astrospheric absorption was used to determine stellar mass-loss rates ($\dot{M}_\star$). However, it is crucial to note that this absorption primarily constrains not the mass-loss rate itself, but rather the wind's dynamic pressure ($P \propto \dot{M}_\star\cdot V_0$), as this pressure governs the spatial extent of the astrosphere. The characteristic size of the interaction region scales approximately with the square root of the ratio between the wind's dynamic pressure and the pressure of the interstellar medium \citep[see equation~\ref{bk1} and][]{Baranov1970, 1971Baranov}. Consequently, higher dynamic pressure results in a larger astrosphere, which increases the integrated column density of neutrals and thus strengthens the absorption signature.} 

\textcolor{orange}{In Wood's studies, a common assumption was that the terminal velocity of solar-like stellar winds ($V_0$) could be approximated by the solar value of $\sim400$ km/s. As \cite{Wood2002} explicitly noted, this assumption may introduce significant systematic uncertainty in the derived mass-loss rates. Since the fundamental constrained parameter is the wind's dynamic pressure, Wood concluded that the inferred mass-loss rate must be inversely proportional to the assumed wind velocity ($\dot{M}_\star \propto V_0^{-1}$). For instance, if a star's actual wind speed is half the solar value, the derived $\dot{M}_\star$ would be twice the value obtained under the standard assumption.}

\textcolor{orange}{Wood's scalability hypothesis inherently presumes that variations in wind speed affect only the overall strength of the absorption without altering the shape or specific features of the spectral profile. However, this fundamental assumption remains largely untested, raising critical questions about spectral morphology: could astrospheric absorption spectra from stars with non-solar wind velocities represent simply scaled versions of heliospheric spectra, or might they exhibit qualitatively different absorption characteristics?}


In this work, we address the question through a parametric study in the frame of a two-dimensional axisymmetric kinetic-hydrodynamic model of an astrosphere. We systematically vary the stellar wind terminal velocity ($V_0$) while maintaining a constant dynamic pressure ($\dot{M}_\star\cdot V_0 = \mathrm{const}$). This ensures the overall size of the astrosphere remains approximately unchanged. 



The structure of this paper is as follows. In Section~\ref{M}, we describe the governing equations, the kinetic-hydrodynamic model of the astrosphere, and the adopted boundary conditions and parameters for the parametric study (Subection~\ref{bound}). Subsection~\ref{Numeric} outlines the numerical methodology, detailing the coupling between the fluid and kinetic solvers and the specific numerical schemes employed. In Section~\ref{result}, we present and discuss our key results: the distributions of different hydrogen populations, the global structure of the astrosphere for various stellar winds, and the analysis of the synthetic Lyman-$\alpha$ absorption spectra that reveals the neutral wind signature. Finally, Section~\ref{concl} summarises the main conclusions of our study. \textcolor{orange}{\ref{dim} presents the dimensionless formulation of the problem and highlights the place of the current work within the context of previous research.}

\section{Model}

\label{M}

The present study employs a two-dimensional axisymmetric model of an astrosphere, which considers the interstellar medium as a two-component mixture consisting of neutral atomic hydrogen gas and a quasi-neutral electron-proton plasma. The plasma is described by the equation of state $p_\mathrm{p} = 2 n_\mathrm{p} k_\mathrm{B} T_\mathrm{p}$ ($k_\mathrm{B}$ denotes the Boltzmann constant), and its dynamics is governed by the Euler equations for a perfect monoatomic gas with constant heat capacities ($\gamma = 5/3$):

\begin{align}
\begin{cases}
\dfrac{\partial \rho_\mathrm{p}}{\partial t} + \textbf{div}( \rho_\mathrm{p}{\bf V}_\mathrm{p}) = Q_1,\\[3mm]
\dfrac{\partial (\rho_\mathrm{p} {\bf{V}}_\mathrm{p})}{\partial t} + \textbf{div}( \rho_\mathrm{p}{\bf V}_\mathrm{p} {\bf V}_\mathrm{p} + p_\mathrm{p}\hat{\bf I}) = \bf Q_2,\\[3mm]
\dfrac{\partial E_\mathrm{p}}{\partial t} + \textbf{div}( (E_\mathrm{p}+p_\mathrm{p}) {\bf V}_\mathrm{p}) = Q_3 + Q_{3,\mathrm{e}},\\[3mm]
\end{cases}
\label{sys1}
\end{align}

where
$
E_\mathrm{p} = \frac{p_\mathrm{p}}{\gamma-1}+ \frac{\rho_\mathrm{p} V_\mathrm{p}^2}{2}
$ is the total energy density of the plasma; $\rho_\mathrm{p},\ p_\mathrm{p},\  {\bf{V}}_\mathrm{p}$ denote the plasma density, the thermal pressure, and the velocity vector, respectively. The sources $Q_1,\ \mathbf{Q_2},\  Q_3,\  Q_{3,\mathrm{e}}$ represent the mass, momentum, and energy gained by the plasma due to charge exchange with hydrogen atoms and photoionization; their expressions are provided below.

The interaction between the neutral component and the plasma is incorporated via source terms in the momentum and energy equations. In contrast, interstellar hydrogen is treated within a kinetic framework. \textcolor{orange}{This approach is required due to the large mean free paths of hydrogen atoms, which are comparable to the size of the astrosphere itself. Consequently, the velocity distribution function becomes strongly non-Maxwellian within the interaction region, precluding an accurate fluid description \citep[see, e.g.][]{izmod2001}. The hydrogen behaviour is therefore described by solving the Boltzmann kinetic equation for the velocity distribution function $f_\mathrm{H}$:}

\begin{align}
	&\frac{\partial f_\mathrm{H}}{\partial t} + {\bf v_\mathrm{H}}\cdot \dfrac{\partial f_\mathrm{H}}{\partial \bf r} + \dfrac{\bf F}{m_\mathrm{H}} \cdot \dfrac{\partial f_\mathrm{H}}{\partial {\bf v_\mathrm{H}}} =  \notag \\
	 &= -f_\mathrm{H}\ n_\mathrm{p} \int |{\bf v}_\mathrm{H} - {\bf v}_\mathrm{p}|\ \sigma_{\mathrm{ex}}(|{\bf v}_\mathrm{H} - {\bf v}_\mathrm{p}|)\ f_\mathrm{p}({\bf v}_\mathrm{p})\ d{\bf v}_\mathrm{p}\ + \label{kma} \\
	&+ f_\mathrm{p}({\bf v_\mathrm{H}})\ n_\mathrm{p} \int |{\bf v}^*_\mathrm{H} - {\bf v}_\mathrm{H}| \sigma_{\mathrm{ex}}(|{\bf v}^*_\mathrm{H} - {\bf v}_\mathrm{H}|) f_\mathrm{H}({\bf v}^*_\mathrm{H}) d{\bf v}^*_\mathrm{H} \notag.
\end{align}


\textcolor{orange}{The vectors ${\bf v}_\mathrm{p}$ and ${\bf v}_\mathrm{H}$ represent the particle velocities of the proton plasma and hydrogen atoms, respectively.} The force {$\bf F$} encompasses the combined effect of stellar gravity and radiation pressure.

The charge exchange cross-section $\sigma_{\mathrm{ex}}(u)$ \textcolor{orange}{(in $\mathrm{cm}^2$)} is given by the following expression, where u is in cm/s \citep{Lindsay2005}:
\begin{align}
\sigma_{\mathrm{ex}}(u) = \left(2.2835 \cdot 10^{-7} - 1.062 \cdot 10^{-8} \mathrm{ln}(u) \right)^2
\end{align}

Here and in what follows, the influence of the force {$\bf F$} on the global flow pattern is assumed to be negligible. This simplification implies that hydrogen atoms travel along rectilinear trajectories between charge-exchange collisions with protons. While trajectory deflection due to {$\bf F$} becomes significant near the star (within several AU, see  \cite{izmod2015}), this effect is confined to a scale that is negligible for the global problem, which has a characteristic size of several hundred AU.

The velocity distribution function of plasma protons $f_\mathrm{p}$ is assumed to be locally Maxwellian:
\begin{align}
f_\mathrm{p}({\bf v}_\mathrm{p}) & = (\sqrt{\pi} c_\mathrm{p})^{-3} \mathrm{exp}\left( -\dfrac{ ({\bf v}_\mathrm{p} - {\bf V}_\mathrm{p})^2}{c_\mathrm{p}^2}\right),\\
c_\mathrm{p} & = \sqrt{\dfrac{2k_\mathrm{B} T_\mathrm{p}}{m_\mathrm{p}}}, \nonumber
\end{align}
where $T_\mathrm{p}$ denote plasma temperature.

The expressions for the sources of mass, momentum and energy in plasma (see System~\ref{sys1}) can be written as follows:
\begin{align}
\begin{cases}
{Q}_1 & = m_\mathrm{p}\cdot n_\mathrm{H}\cdot \nu_{\mathrm{ph}},\\
{\bf Q}_2 & = m_\mathrm{p}\cdot \int \ \nu_{\mathrm{ph}}\cdot {\bf v}_\mathrm{H} \cdot f_\mathrm{H}({\bf v}_\mathrm{H})\ d{\bf v}_\mathrm{H}\ + \\ & n_\mathrm{H}\cdot \rho_\mathrm{p}\cdot \iint |{\bf v}_\mathrm{p} - {\bf v}_\mathrm{H}| \cdot \sigma_{\mathrm{ex}}(|{\bf v}_\mathrm{p} - {\bf v}_\mathrm{H}|) \cdot ({\bf v}_\mathrm{p} - {\bf v}_\mathrm{H}) \\
&\cdot f_\mathrm{H}({\bf v}_\mathrm{H})\cdot f_\mathrm{p}({\bf v}_\mathrm{p})\ d{\bf v}_\mathrm{H} d{\bf v}_\mathrm{p},\\
Q_3 & = m_\mathrm{p}\cdot \int \ \nu_{\mathrm{ph}}\cdot {\bf v}^2_\mathrm{H}/2 \cdot f_\mathrm{H}({\bf v}_\mathrm{H})\ d{\bf v}_\mathrm{H}\ + \\ & n_\mathrm{H}\cdot \rho_\mathrm{p}\cdot \iint |{\bf v}_\mathrm{p} - {\bf v}_\mathrm{H}|\cdot \sigma_{\mathrm{ex}}(|{\bf v}_\mathrm{p} - {\bf v}_\mathrm{H}|) \cdot \dfrac{{\bf v}_\mathrm{H}^2 - {\bf v}_\mathrm{p}^2}{2} \\
&\cdot f_\mathrm{H}({\bf v}_\mathrm{H}) \cdot f_\mathrm{p}({\bf v}_\mathrm{p})\ d{\bf v}_\mathrm{H} d{\bf v}_\mathrm{p}, \\
{Q}_{3,\mathrm{e}} & = n_\mathrm{H}\cdot \nu_{\mathrm{ph}}\cdot E_{\mathrm{ph}},
\end{cases}
\label{syssource}
\end{align}

where $\nu_{\mathrm{ph}} = 1.67\cdot 10^{-7}(R_\mathrm{E}/R)^2\ \mathrm{s}^{-1}$ ($R_\mathrm{E}$ = 1 AU) is the photoionization rate \citep{izmod2015}, $E_{\mathrm{ph}}$ is the mean photoionization energy (4.8 eV).



\subsection{Boundary conditions}
\label{bound}
The boundary conditions for the plasma and neutral components are defined as follows. A coordinate system is adopted with its origin at the star and the \textit{X}-axis aligned with the upwind direction.

In the case of fully ionised interstellar medium, or when the effect of atoms can be neglected, the size of an astrosphere is determined by the ratio of the dynamic pressure of the stellar wind to the pressure of the interstellar medium \citep{Baranov1970, 1971Baranov}: 
\begin{align}
L \sim \sqrt{ \left({\dot{M}_\star V_0}\right)/\left({4 \pi \rho_{\mathrm{p},\infty} V^2_{\infty}}\right)}, \label{bk1}
\end{align}
where $\dot{M}_\star = 4\pi\rho V_0 R^2$ is the stellar mass loss rate and $V_0$ is the terminal wind velocity. For two-component models, the effect of charge change between components can reduce the distance by a factor of two \citep[see, e.g., Figure~2 in][]{Korolkov2024A}. The objective of this work is to investigate astrospheres with a size approximately equal to that of the heliosphere but with different combinations of stellar wind density and velocity. This approach is equivalent to maintaining a constant stellar wind dynamic pressure. Within this framework, we vary the value of $V_0$, thereby modelling stellar winds of different densities and velocities (see Table~\ref{tab:wind_parameters}). 

\begin{table}[h!]
\centering
\caption{Parameters of the stellar wind models. $\dot{M}_\star \cdot V_0$ remains constant, the same as in the solar wind. \textcolor{orange}{The values presented are determined by dimensionless parameters (see \ref{dim}).}}
\begin{tabular}{|c|c|c|}
\hline
Model star & $\dot{M}_\star\ (\dot{M}_\odot)$ & $V_0$ (km/s)\\
\hline
$\mathrm{M}_{426}$ (Sun) & 1 & 426 \\
\hline
$\mathrm{M}_{311}$ & 1.37 & 311 \\ 
\hline
$\mathrm{M}_{208}$ & 2.05 & 208\\ 
\hline
$\mathrm{M}_{104}$ & 4.1 & 104\\ 
\hline
$\mathrm{M}_{78}$ & 5.47 & 78\\ 
\hline
\end{tabular}
\label{tab:wind_parameters}
\end{table}


The interstellar medium is represented as a parallel plasma flow with specified conditions: density $n_{\mathrm{p},\infty} = 0.04\ \mathrm{cm}^{-3}$, velocity $V_{\infty} = 26.4\ \mathrm{km/s}$, and temperature $T_{\infty} = 6530\ \mathrm{K}$. \textcolor{orange}{These values are chosen to be close to the parameters of the interstellar medium surrounding the heliosphere.}
 
We assume a Maxwellian velocity distribution function for the interstellar hydrogen atoms at the boundary:
 \begin{align}
& f_{\mathrm{H}, \infty}({\bf V}_\mathrm{H})  = (\sqrt{\pi} c_\infty)^{-3} \mathrm{exp}\left( -\dfrac{ ({\bf V}_\mathrm{H} - {\bf V}_\infty)^2}{c_\infty^2}\right),\\
& c_\infty  = \sqrt{\dfrac{2k_\mathrm{B} T_\infty}{m_\mathrm{p}}}, \nonumber
\end{align}

where $T_\infty,\ {\bf V}_\infty$ denote temperature and velocity of the unperturbed local interstellar medium \textcolor{orange}{(here it is assumed that hydrogen and plasma in the interstellar medium are in local thermodynamic equilibrium, sharing identical temperatures and velocities).} The hydrogen number density is \textcolor{orange}{ $n_{\mathrm{H},\infty} = 0.12\ \mathrm{cm}^{-3}$}.

\begin{figure*}
\includegraphics[width=\textwidth]{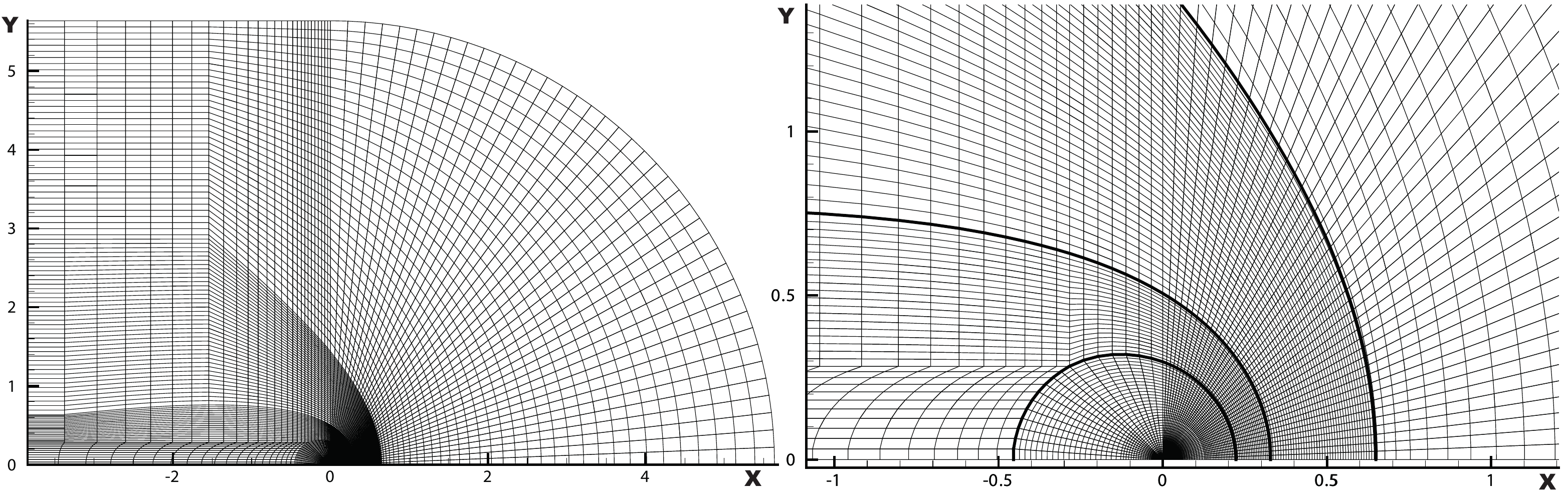}
\caption{
\textcolor{orange}{Examples of the computational grids used, with the capability to capture the main discontinuities. The left panel shows the entire flow domain, while the right panel focuses on the head region. Discontinuity surfaces are marked with black lines in the right panel.}
}
\label{fig:setka}
\end{figure*}

\subsection{Numerical approach}
\label{Numeric}

\begin{figure*}
\includegraphics[width=\textwidth]{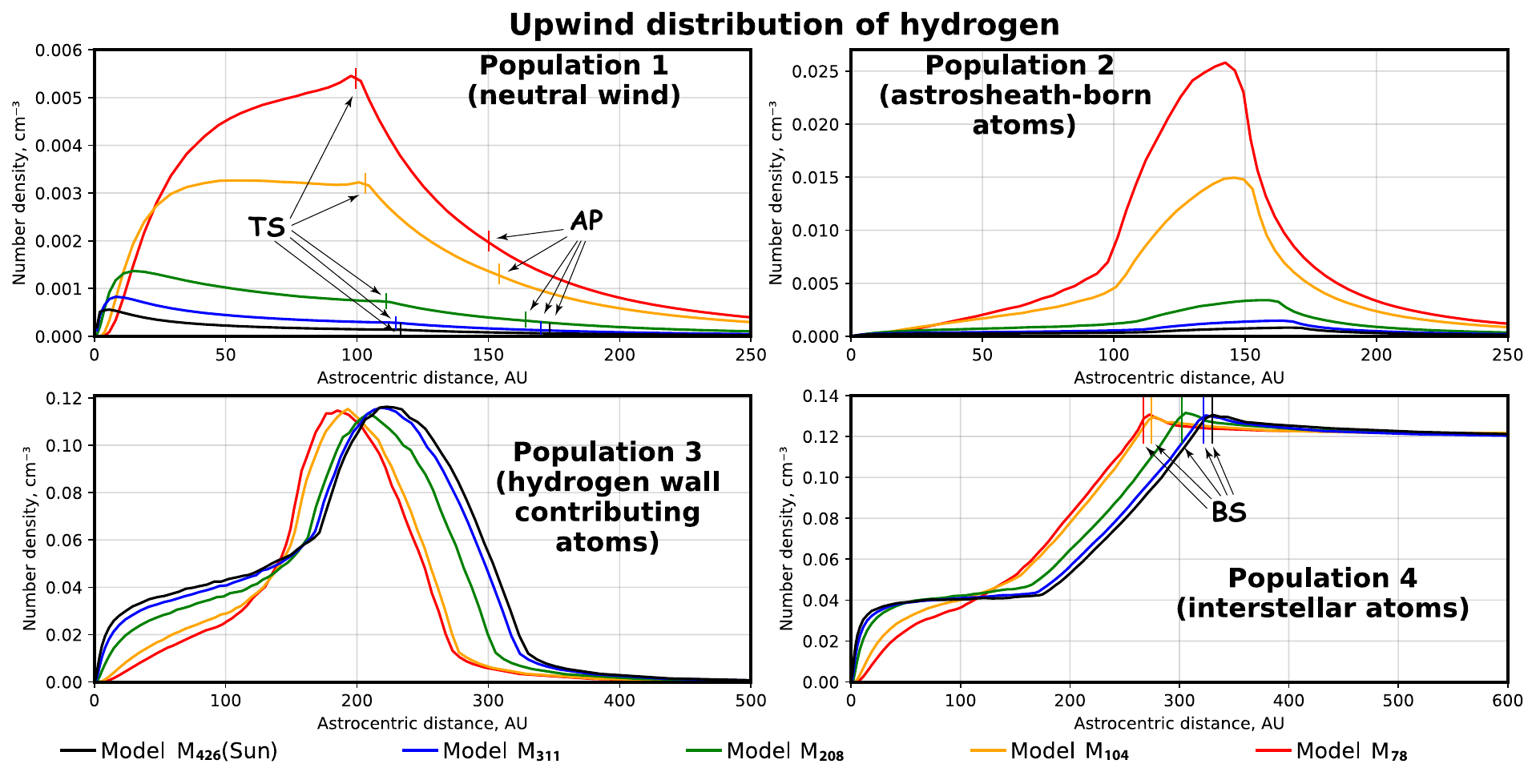}
\caption{
Number densities of the four hydrogen atom populations along the upwind direction. Model notations are shown in Table~\ref{tab:wind_parameters}. \textcolor{orange}{The arrows indicate the position of the plasma discontinuity surfaces (TS - the termination shock, AP - the astropause, BS - the bow shock). }
}
\label{fig:hydro}
\end{figure*}


This subsection outlines the numerical methodology for this study. \textcolor{orange}{The solution strategy couples two distinct physical models, a fluid description for the plasma and a kinetic treatment for the neutrals, within a global iterative scheme. The process is initialised by solving the system of fluid equations~(\ref{sys1}) for the plasma in the absence of neutral atoms, i.e., with the momentum and energy source terms set to zero. On each iteration, the plasma equations are solved using a time relaxation method until a steady state solution is achieved. In the subsequent step, these plasma parameters are used to solve the kinetic equation~(\ref{kma}) via the Monte Carlo method, which yields the corresponding neutral atom distribution. This distribution, in turn, is used to compute the source terms for the fluid equations. The cycle, solving the fluid equations with the current source terms, then updating the neutral distribution via the kinetic solution, is repeated until a global, self-consistent solution is achieved, with both the plasma and neutral atom distributions converging to a steady state. A steady state for the plasma is considered reached when its parameters cease to evolve in time, meaning their numerical change between time steps becomes negligible. Typically, 7–10 global iterations are sufficient for the plasma and neutral atom distributions to fully converge to a final steady state.}
Further details on the Monte Carlo implementation are available in \cite{Sobol}, while the overarching algorithm, central to the Moscow model of the heliosphere, is detailed in \cite{Malama1991}. 

\textcolor{orange}{Note that due to cylindrical symmetry, the Euler equation (\ref{sys1}) is two-dimensional in space, whereas the kinetic equation (\ref{kma}) is five-dimensional (two spatial dimensions and three velocity dimensions). In the Monte Carlo method, it is essential to track the three-dimensional trajectories of particles through the astrosphere, as in the absence of force $\bf F$, these trajectories remain rectilinear until charge exchange occurs. However, the projection of a three-dimensional rectilinear trajectory onto the cylindrical (x, r)-plane generally does not preserve its rectilinearity.}

To solve (\ref{sys1}), we employ the finite volume method as detailed in \cite{godunov1976}, utilising a time relaxation approach. The Riemann problem at cell interfaces is handled using either the approximate HLLC solver from \cite{Miyoshi} or the exact solver from \cite{godunov1976}. Spatial reconstruction within cells is achieved through the minmod limiter \citep{1990Hirsch}, while the source terms ${\bf Q}_2$ and $Q_3$ are computed using the Monte Carlo code.

All calculations are performed on a two-dimensional computational grid designed to fit the key discontinuity surfaces: the termination shock, the astropause, and the bow shock (see Figure~\ref{fig:setka}). \textcolor{orange}{The model employs a numerical grid of approximately 15000 cells. In the head region, the grid has a spherical structure. The radial distribution of cells is structured across distinct flow regions: the inner shock layer is discretised with 30 cells, while both the outer shock layer and the region of supersonic stellar wind are resolved with 40 cells each. A total of 35 cells span the interstellar medium region, with grid clustering towards the outer shock front. In the angular direction, the grid consists of 100 rays uniformly distributed from 0 to $\pi$. To verify the numerical convergence of the results, supplementary test calculations were conducted on a refined grid with twice the resolution in every direction.}

\textcolor{orange}{The calculation is performed in a cylindrical coordinate system where the X-axis is oriented in the upwind direction. The Y-axis represents the perpendicular radial distance r. The cylindrical symmetry is accounted for within the Euler equations through the inclusion of corresponding geometric source terms, implemented according to standard methodology \citep[e.g., ][]{godunov1976}.}

For plasma parameters, the so-called rigid boundary conditions are applied at the inlet boundaries, where density, velocity, and pressure are held fixed, while soft boundary conditions imposing zero gradients for all parameters are used at the outlet boundaries. Numerical tests have confirmed that these boundary conditions do not introduce spurious disturbances into the flow.

The calculation of synthetic Lyman-$\alpha$ absorption spectra is integrated into the Monte Carlo solution of the kinetic equation. During this process, the projection of the velocity distribution function onto the specified line of sight is accumulated within each computational cell \textcolor{orange}{\citep[this projection is evaluated as the statistical average over a large ensemble of Monte Carlo test particles, see, e.g.,][]{Malama1991}}. This accumulated projection then serves as the basis for computing the absorption profile. 

\section{Results and discussion}
\label{result}

\begin{figure}
\includegraphics[width=0.48\textwidth]{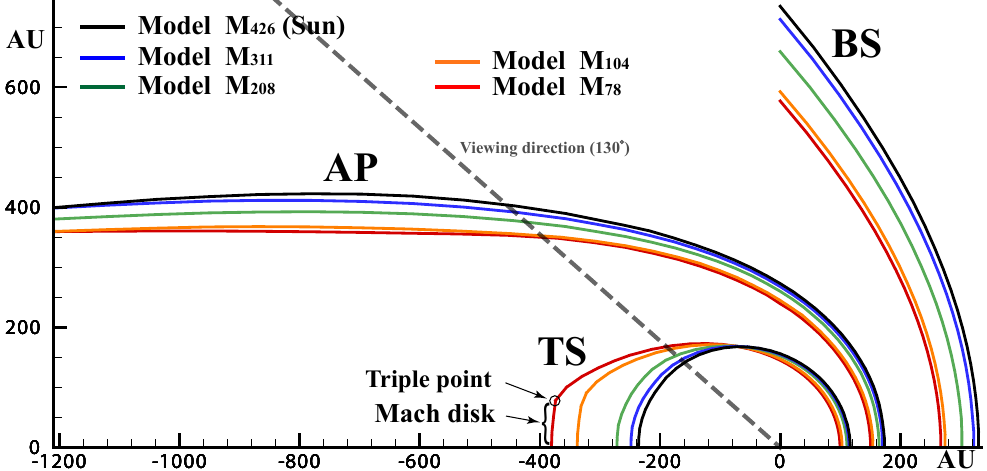}
\caption{
Locations of the discontinuity surfaces (TS - the termination shock, AP - the astropause, BS - the bow shock) for different stellar wind models. The gray dotted line indicates the line of sight toward the astrospheric tail ($130\degree$).
}
\label{fig:surf}
\end{figure}

\begin{figure*}
\includegraphics[width=\textwidth]{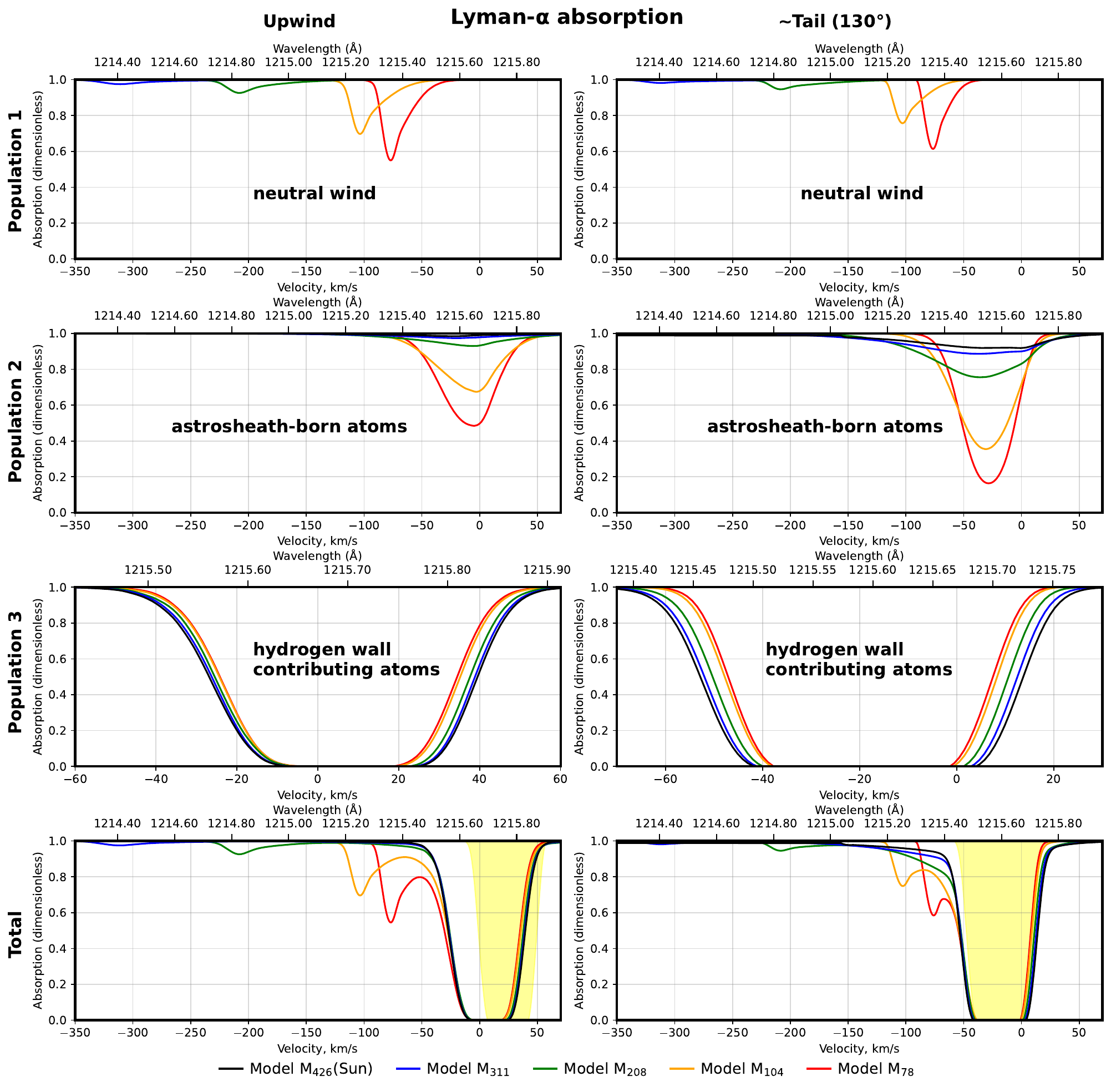}
\caption{
Dimensionless synthetic Lyman-$\alpha$ absorption spectra for the upwind and tail ($130\degree$) lines of sight through the model astrosphere. Columns correspond to different lines of sight, while rows show contributions from different atom populations (1-3) and the total absorption. \textcolor{orange}{The yellow shaded area indicates the approximate combined absorption from population 4 and unperturbed interstellar hydrogen, which is excluded from the total spectra. All spectra are shown in the Solar rest frame.}}
\label{fig:parametric}
\end{figure*}

Figure~\ref{fig:hydro} demonstrates the hydrogen atom number densities as a function of the astrocentric distance (in AU). The distributions are shown for the upwind direction and for models from Table~\ref{tab:wind_parameters}. The four panels display distributions of the four distinct atom populations. 
\textcolor{orange}{These populations are defined by the location where the atoms are created (the Monte Carlo method can track the origin of each simulated test particle).} They are categorised according to the methodology of \cite{Izmodenov2000} and are used here solely for the interpretation of the results.\textcolor{orange}{ This categorisation is physically meaningful because atoms born within a specific plasma region inherit the bulk properties (velocity, temperature) of their parent protons at that location, leading to significantly different characteristics for each population.} The four populations are: (1) atoms created by charge exchange between interstellar hydrogen and supersonic stellar wind protons, referred to as the neutral wind or population 1; (2) atoms originating in the inner astrosheath, where the stellar wind is heated and decelerated after crossing the termination shock (population 2 or astrosheath-born atoms); (3) atoms born in the outer astrosheath from the heated and slowed-down interstellar medium plasma, which constitute the so-called hydrogen wall (population 3); and finally, (4) pristine interstellar hydrogen atoms entering the computational domain from its upstream boundary, as well as \textcolor{orange}{those generated by charge exchange that occurs in the supersonic interstellar flow upstream of the bow shock (population 4). Population 4 retains properties closest to those of the unperturbed interstellar hydrogen.}


Figure~\ref{fig:hydro} demonstrates that the smaller the terminal velocity in the model, the higher the number densities of populations 1 and 2 in the models. 
This effect has a simple explanation. The source of populations 1 and 2 depends on the charge exchange frequency, which, in turn, linearly depends on the number density of protons. Since in this study we assume the dynamic pressure of the stellar wind to be the same in all models ($\rho V^2 = \mathrm{const}$), the density is inversely proportional to the square of the terminal velocity. Therefore, the smaller the terminal velocity, the larger the proton number density in the stellar wind, and the charge exchange frequency, which results in a higher source of the populations.


The distributions of populations 3 and 4 shown in Figure~\ref{fig:hydro} are quite similar for all models. This is expected, as the parameters of the interstellar medium remain unchanged. However, the high efficiency of charge exchange in the models with slow and dense stellar winds (models $\mathrm{M}_{104}$ and $\mathrm{M}_{78}$)  is also seen in the slopes of the density profiles inside the astropause, which is approximately at 150 AU. For Models $\mathrm{M}_{104}$ and $\mathrm{M}_{78}$, these slopes are significantly steeper, whereas for Models $\mathrm{M}_{426}$ and $\mathrm{M}_{311}$, Population 4 atoms within the astrosphere ($50 \lesssim X \lesssim 180$ AU), the gradient is nearly zero, which means that charge exchange is ineffective in the inner astrosheathes in these models.

Figure~\ref{fig:surf} shows the locations of the discontinuity surfaces (the termination shock, the heliopause and the bow shock) for the models from Table~\ref{tab:wind_parameters}. The positions of the discontinuities in the upwind region are consistent with the previous conclusion on the more efficient charge exchange process for the models with smaller terminal velocities.
For such models, the astropause and both shocks are located closer to the star since the larger momentum is transferred from the interstellar hydrogen to the plasma in these models.

Another interesting result is the shape of the termination shock in the tail region. For the models with small terminal velocities, the termination shock becomes elongated in the tailward direction. For Model $\mathrm{M}_{78}$, its shape exhibits a kink, forming a triple point and a Mach disk. Such a termination shock configuration is characteristic of the pure gas-dynamic case \textcolor{orange}{(i.e., a model in which the interstellar neutral hydrogen density is set to zero, thus removing its influence) and was previously obtained in gas-dynamic astrospheric models (for the first time for the heliosphere in \citealtc{baranov1993}, Figure~2; then, for example, \citealtc{Korolkov2020}, \citealtc{Korolkov2023}).} In the case of Models $\mathrm{M}_{104}$ \textcolor{orange}{and $\mathrm{M}_{78}$}, the dense wind effectively "filters" the interstellar hydrogen, preventing it from penetrating the tail region. As a result, the solution in this region closely resembles the gas-dynamic case with no neutrals. \textcolor{orange}{The Mach disk can also arise in other astrophysical contexts, such as the interaction of a stellar or solar wind with a cometary body \citep{Wallis1976}.}

Figure~\ref{fig:parametric} presents synthetic Lyman-$\alpha$ absorption spectra in dimensionless form \citep[stellar profiles $I_0(\lambda) = 1$, see ][]{Izmodenov2002}. \textcolor{orange}{For the direct convenience of comparison with actual observations, all spectra are shown in the Solar rest frame, that is, as they would be recorded by the Hubble Space Telescope, under the assumption of zero stellar radial velocity relative to the Sun; otherwise, the entire spectrum would need to be shifted by the corresponding projection of this velocity. The left and right columns correspond to two different lines of sight through the model astrosphere: the upwind direction and the tail region observed at $130\degree$ from the upwind axis (see Figure~\ref{fig:surf}, grey dotted line), respectively.} Each row in the Figure represents absorption spectra produced by different populations (1-3) of astrospheric H atoms. The total absorption spectra are shown in the bottom row of the spectra. \textcolor{orange}{The synthetic spectra in Figure~\ref{fig:parametric} are calculated excluding absorption from population 4 and the unperturbed interstellar hydrogen along the line of sight. The yellow shaded region represents an estimate of this combined absorption. It is evident that interstellar hydrogen determines the redward (right) part of the spectrum. Therefore, our current analysis of astrospheric absorption is focused on the blueward (left) side of the profile, which is not obscured by the interstellar hydrogen. We note that this shaded area is approximate; the exact shape and depth of the interstellar absorption depend on the specific line-of-sight column density to the star and the fundamental parameters of the local interstellar medium (e.g., its temperature and velocity) and could be somewhat broader.} 



The major difference in the spectra obtained with different models \textcolor{orange}{stems from} the absorption produced by populations 1 and 2. The centre of the absorption line produced by population 1 corresponds to the terminal velocity of the stellar wind \textcolor{orange}{(this velocity is negative, due to the astrospheric wind moving toward the observer)}. For Models $\mathrm{M}_{208}$-$\mathrm{M}_{78}$, this absorption is sufficiently pronounced to be detectable in the total spectrum. This fact can be directly utilised to identify dense stellar winds with velocities up to $\sim$200 km/s. For higher velocity winds, detecting such absorption becomes challenging. For model $\mathrm{M}_{426}$, this absorption does not exceed 1\%, and \textcolor{orange}{also this absorption may be Doppler shifted beyond the wing of the stellar Ly $\alpha$ emission profile.}

Absorption produced by population 2  is also discernible in the total spectra. It is particularly evident in the tail direction, manifesting as a slope in the spectrum at velocities of –100 to –50 km/s. Although this observational signature is more subtle, its specific localisation in the –100 to –50 km/s velocity range means that, given sufficiently high spectral resolution, it could provide information about stellar winds with velocities up to $\sim$300 km/s. This significantly expands the potential application range of this diagnostic method.


\textcolor{orange}{Based on Figure~\ref{fig:parametric}, we can conclude the most favourable lines of sight through the astrosphere (i.e., directions defined in the stellar rest frame) for observing Lyman-$\alpha$ absorption. For detecting the hydrogen wall (population 3), upwind or near-upwind directions are preferable, as the plasma in the outer astrosheath is more heated and decelerated precisely in the upwind direction. This results in better separation and distinguishability between the absorption from the interstellar hydrogen and the hydrogen wall. For observing astrosheath absorption (from population 2), tailward directions are most suitable, since the inner astrosheath is wider in this region (leading to a longer integration column). The observation of neutral wind absorption (population 1) is relatively insensitive to the line-of-sight direction.}

An important observational implication of our results is the potential spectral overlap between the neutral wind absorption (population 1) and the interstellar Deuterium (D I) line. The D I line has a fixed isotopic shift of $\approx -82$ km/s from the main H I line.  As it is seen from our calculations for models $\mathrm{M}_{104}$ and $\mathrm{M}_{78}$, the absorption produced by population 1 may directly overlap with  D I line. This makes the analysis of the observed spectra quite challenging. The absorption produced by population 1 could be mistakenly considered as the interstellar D I absorption or blended with it. Such misidentification would potentially lead to an overestimation of the deuterium column density, systematically affecting the derived D/H ratio -- a critical parameter for cosmological and Galactic chemical evolution studies. Therefore, when analysing Ly $\alpha$ spectra of the stars with small terminal velocities, it is essential to use detailed astrospheric models to disentangle the contribution of the neutral stellar wind from that of interstellar deuterium.

\textcolor{orange}{Note that the studies by \cite{Wood2005} did not reveal definitive signatures of absorption by the neutral stellar wind (population 1). However, it should be emphasised that the authors likely did not account for the potential presence of such absorption. In particular, explaining certain spectral features required invoking multiple deuterium populations \citep[see Figure 3 in][]{Wood2005}, with these absorptions being significantly separated for some stars (e.g., HD 32008). As noted in the previous paragraph, the neutral wind could naturally account for such features. Furthermore, as acknowledged by the authors, many spectra exhibit substantial noise, and some spectral features interpreted merely as noise could potentially indicate neutral wind absorption \citep[e.g., the absorption at approximately 1215.1 $\mathring{\mathrm{A}}$ in the spectrum of HD 128987 in Figure 7][]{Wood2005}.}

\textcolor{orange}{This work is primarily applicable to stars with slower stellar winds, where the neutral wind component can be clearly distinguished in spectra. Promising candidates include red dwarfs  \citep[e.g., the stellar wind speed of Gliese 436 is 85 km/s][]{Bourrier2016} and orange dwarfs \citep[e.g., HD 189733 has a wind speed of 190 km/s ][]{Bourrier2013}. Evolved stars, such as red giants/supergiants, and some blue supergiants, also represent excellent candidates for detecting absorption from neutral winds.}

\textcolor{orange}{For solar-type stars, the direct observation of neutral wind absorption is generally beyond the method's applicability limit, as their wind speeds are presumed to be similar to the solar wind ($\sim$400 km/s). However, potential latitudinal variations in the wind structure could offer some prospects. If the wind speed at a star's low latitudes is close to 200 km/s and the line of sight aligns with this direction, the conditions might approach the method's upper applicability limit. Nevertheless, for solar-type stars, a second diagnostic capability of the method should not be overlooked: detecting the spectral slope change due to absorption from population 2 (associated with the astrosheath absorption in the tail direction). This effect is more subtle and consequently imposes stricter requirements on the quality of the observed spectra.}

\textcolor{orange}{It should be noted that, for a real stellar wind, the terminal velocity is not a single-valued parameter but is subject to temporal variations. Such variability could be averaged out over time in observations, potentially 'smearing' the resulting absorption signature more than predicted by our steady-state model. Consequently, the observed absorption feature could be broader and less pronounced than in our synthetic profiles. This effect represents a natural limitation of the current modelling approach and could be a subject for future, more detailed studies in the frame of time-dependent models.}

\section{Conclusion}
\label{concl}
In this paper, we present results of the parametric study of the astrospheres and their Lyman-$\alpha$ absorption spectra by varying the terminal velocity of the star and keeping the stellar wind dynamic pressure (${\dot{M}_\star \cdot V_0 = \mathrm{const}}$) (see Table~\ref{tab:wind_parameters}). The dynamic pressure was chosen so that the size of the stellar astrospheres is approximately equal to the heliosphere. The key results can be summarised as follows:

\begin{itemize}
\item The number densities of the heliospheric populations of H atoms - neutral stellar wind (population 1 in our notation) and hot astrosheath neutrals (population 2)  strongly depend on the velocity of the stellar wind. The smaller the velocity,  the larger the number densities of these populations. The astrospheric discontinuities (the termination shock, the heliopause, the bow shock) become closer to the star for the models with the smaller stellar wind velocities.

\item For slow and dense winds (model $\mathrm{M}_{78}$), the shape of the Termination Shock in the tail region becomes qualitatively similar to the shape of the shock in the pure gas-dynamic model when interstellar neutrals are not taken into account. Specifically, a triple point and a Mach disk appeared. This is explained by the effective \textcolor{orange}{"shielding" from} interstellar hydrogen by the dense stellar wind.

\item Analysis of the synthetic Lyman-$\alpha$ absorption spectra produced by our models shows that the absorption produced by population 1 (i.e. by neutral wind) is observable for the stellar winds with velocities up to ~150-200 km/s. More subtle spectral features associated with population 2 (hot astrosheath atoms) suggest a potential method for detecting winds with velocities up to ~300 km/s, substantially expanding the diagnostic range of this technique.

\item We demonstrate that the absorption features produced by the neutral stellar wind (population 1) can potentially be blended with the absorption produced by interstellar deuterium. 

\textcolor{orange}{Future investigations will extend this work to systematically study the influence of other fundamental astrospheric parameters (see \ref{dim}) on the Ly $\alpha$ absorption signature. Of particular interest are regimes with a high interstellar neutral hydrogen density, which could enhance absorption, as may be relevant for stars like HD~61005 \citep{Hines2007}, and regimes with a low relative interstellar velocity, as may be relevant for stars like HD~128987 \citep{Wood2005ApJ}.}

\end{itemize}

\section*{Acknowledgments}
The work was supported by the Russian Science Foundation grant № 25-12-00240, \url{https://rscf.ru/project/25-12-00240/}. 



\bibliographystyle{jasr-model5-names}
\biboptions{authoryear}
\bibliography{bibliography}



\appendix

\section{Dimensionless formulation}
\label{dim}

\textcolor{orange}{This appendix outlines a clear pathway for future research. Its primary goal is to develop a systematic understanding of how the Ly $\alpha$ absorption spectrum depends on each governing parameter of the astrosphere. Such knowledge would not only allow one, in principle, to recover the stellar-wind and local interstellar medium parameters of an arbitrary astrosphere directly from an observed spectrum but would also help answer important open questions regarding the accuracy and uniqueness of such a procedure. The ideal tool for this task would be a comprehensive database of synthetic spectra, computed for a grid of governing parameters.}

\textcolor{orange}{To investigate an arbitrary astrosphere, it is methodologically preferable to formulate the problem in a dimensionless form, thereby reducing the number of independent parameters. In this framework, the absorption spectrum should be expressed in dimensionless units (e.g., as a function of dimensionless velocity or wavelength). For any specific astrosphere, one could then take the pre-computed spectrum corresponding to its set of dimensionless parameters and convert it back into observable, physical units. However, in the present work, for the direct convenience of observational astrophysicists, we present all spectra in dimensional form.}

\textcolor{orange}{We can reformulate the problem in dimensionless form by reducing the number of parameters to four. Let us relate all distances to $L^* = \sqrt{ \left({\dot{M}_\star V_0}\right)/\left({4 \pi \rho_{\mathrm{p},\infty} c^2_{\infty}}\right)}$, all velocities to the thermal velocity $c_\infty$, the plasma density to $\rho_{\mathrm{p},\infty}$, the atom number density to $n_{\mathrm{H}, \infty}$.} 

\textcolor{orange}{As a result, four dimensionless parameters of the problem are obtained:
$$\chi,\ \eta,\ M_\infty,\  \mathrm{Kn}_\infty.$$
Descriptions of these parameters are as follows: \\
(1) $\chi = \dfrac{V_0}{c_\infty}$ is the ratio of the terminal velocity of the star to the thermal velocity of the unperturbed interstellar medium. }
\\
\textcolor{orange}{(2) $\eta = \dfrac{n_{\mathrm{H}, \infty}}{n_{\mathrm{p}, \infty}}$ is the ratio of hydrogen number density to proton number density in the unperturbed interstellar medium.}
\\
\textcolor{orange}{(3) $M_\infty$ is the Mach number of the unperturbed interstellar medium.}
\\
\textcolor{orange}{(4) $\mathrm{Kn}_\infty = \dfrac{l_{\mathrm{ex}, \infty}}{L^*}$ is the Knudsen number, defined as the ratio of the mean free path of an atom in the interstellar medium to the characteristic size of the astrosphere.  The mean free path is calculated as follows:}
\textcolor{orange}{\begin{align}
l_{\mathrm{ex}, \infty} = \dfrac{1}{n_{\mathrm{p}, \infty}\  \sigma_{\mathrm{ex}}(c_\infty)}.
\end{align}}

\textcolor{orange}{Thus, an arbitrary astrosphere within the assumptions of the present work (e.g., in the absence of magnetic fields, etc.) can be characterised by four dimensionless parameters. The Ly $\alpha$ absorption spectra undoubtedly depend on each of these parameters in one way or another. In this work, a parametric study of spectra was conducted with respect to the $\chi$ parameter. In contrast, the other parameters were fixed at values approximately equal to those of the heliosphere ($M_\infty$ $\approx$ 2, $\eta$ $\approx$ 3, $\mathrm{Kn}_\infty$ $\approx$ $0.5$). The model stellar winds $\mathrm{M}_{426}$-$\mathrm{M}_{78}$ from Table~\ref{tab:wind_parameters} correspond to $\chi$ values of 41, 30, 20, 10, and 7.5, respectively.}

\textcolor{orange}{In works \cite{Wood2002, Wood2003, Wood2005ApJ, Wood2014}, authors varied the stellar mass-loss rate $\dot{M}_\star$, or equivalently, the size of the astrosphere $L^*$, thereby investigating absorption spectra as a function of the Knudsen number $\mathrm{Kn}_\infty$. However, this variation was within a relatively small range: the Knudsen number changed by no more than a factor of $\approx$3 across all studies, except for \cite{Wood2014}, where the variation was about a factor of $\approx$22. An investigation of astrospheres over a wide range of Knudsen numbers (6 orders of magnitude) was performed in \cite{Korolkov2024A, Korolkov2024B}, but the authors focused their attention on the plasma and neutral atom flows and did not present absorption spectra.}

\textcolor{orange}{Ly $\alpha$ absorption spectra for six astrosphere models with different $n_\mathrm{H}$ and $n_\mathrm{p}$ are presented in \cite{Izmodenov2002} (this study essentially explores six combinations of the parameters $\eta$ and $\mathrm{Kn}_\infty$). In \cite{Wood2007b}, the behaviour of Ly $\alpha$ absorption spectra was investigated for several models with different magnitudes and directions of the interstellar magnetic field. In this case, the model becomes three-dimensional, introducing two additional dimensionless parameters (magnetic field magnitude and direction - angle from upwind), making a comprehensive parametric study significantly more challenging. A number of other works \citep[see, for example,][]{Wood2007c} present several astrospheres with different sets of defining parameters.}

\textcolor{orange}{Therefore, a systematic and comprehensive parametric study of Lyman-$\alpha$ absorption spectra across the four key dimensionless parameters ($\chi$, $\eta$, $M_\infty$, $\mathrm{Kn}_\infty$) remains an important and relevant task. The creation of a corresponding database of synthetic spectra would facilitate a transition from the analysis of individual cases to a general method for diagnosing the properties of astrospheres from observational data.}



\end{document}